\documentclass[aps,twocolumn,showpacs,superscriptaddress,amsmath]{revtex4}
\usepackage{graphicx}
\usepackage{graphics}
\usepackage{epsfig}
\newcommand{\be}{\begin{equation}}
\newcommand{\ee}{\end{equation}}
\newcommand{\bea}{\begin{eqnarray}}
\newcommand{\eea}{\end{eqnarray}}
\begin{document}

\title{Narrow optical filtering with plasmonic nanoshells}
\author{Y.\ B.\ Martynov}
\affiliation{Institute of Applied Acoustics, 141980 Dubna, Russia}
\author{R.\ G.\ Nazmitdinov}
\affiliation{Departament de F{\'\i}sica,
Universitat de les Illes Balears, E-07122 Palma de Mallorca, Spain}
\affiliation{Bogoliubov Laboratory of Theoretical Physics,
Joint Institute for Nuclear Research, 141980 Dubna, Russia}
\author{I.\ A.\ Tanachev}
\affiliation{Institute of Applied Acoustics, 141980 Dubna, Russia}
\affiliation{Dubna University, 141980 Dubna, Russia}
\author{P.\ P.\ Gladyshev}
\affiliation{Institute of Applied Acoustics, 141980 Dubna, Russia}
\affiliation{Dubna University, 141980 Dubna, Russia}

\begin{abstract}
Narrow optical band pass filters are widely used in systems with optical 
processing of information, color displays development and optical computers.
We show that such ultra filters can be created by means of  nanoparticles which consist of 
a dielectric sphere and a metallic shell. The components can be adjusted such that 
there is a remarkable transparency at the desired 
wavelength range, while a strong absorption takes place outside of this region.
\end{abstract}

\date{\today}
\maketitle

Nowadays, narrow band pass filters to be used in dense
wavelength division multiplexing are among topical problems
in the fiber optics communications field.  Although  the design of these filters is 
relatively straightforward, the classical techniques suffers from 
uncontrolled errors \cite{wil}. For near-infrared astronomical imaging, 
the efficient suppression of some emission lines in the Earth's sky together 
with an optimised transmittance band is a strategic need  
to increase the sensitivity of ground-based instruments for 
a valuable improvement in observing efficiency \cite{ast}.
It is noteworthy that  the advent of nanotechnology gives impetus 
to the field of plasmonics (cf \cite{pol}) which  enables one to operate with 
light at the nanoscale, well below the scale accessible for the classical techniques.
For example, a high sensitivity can be achieved in biosensing technology using 
a plasmonic material \cite{bio}.
Various applications of plasmonics appear
for nanoscale switches \cite{kras}, imaging below the
diffraction limit \cite{pen}, materials with negative refractive index \cite{n1},
to name just a few. It is a challenge to understand how plasmon excitations and 
light localization at nanoscale might be used advantageously for a narrow band pass filtering.

Among well-known narrow band pass optical filters are those that are 
based on the Christiansen effect \cite{chr}.
 This effect is due to scattering by small particles in a transparent medium at a 
wavelength for which the refractive index of the particle material 
and that of the medium are equal. These filters
transmit unscattered light at this wavelength and incoherently scatter light 
of other wavelengths. A change of the transmission behavior 
of this dispersion filter can be achieved by variation of the composed materials
and external conditions. 
A fundamental constraint in manipulating light with such filters is a
strong relationship between the transmission  bandwidth and  
the detector aperture \cite{hof}. 
The major goal of the present paper is to demonstrate that the efficiency of such
filters can be drastically improved by means of small nanoparticles 
with metallic nanoshells. Note that in such nanoparticles the absorption is 
greatly enhanced in contrast to the scattering at plasmon resonance frequency.

A particular interesting object is a nanoparticle composed of a dielectric core 
and  a homogenous metallic nanoshell. Mie scattering theory predicts that by 
varying the ratio of the shell thickness with the respect to the overall diameter 
of the particle it is possible to obtain the invisibility of the nanoparticle at 
a specific wavelength \cite{ker,boh,alu}. In this case a scattering cancellation 
is based on the negative local polarizability of a metallic nanoshell 
with  respect to the positive dielectric core polarizability.
It was shown that the dipolar term is dominant in the Mie expansion 
for light scattering from a spherical small plasmonic particle 
with  a radius $a\leq\lambda/10$, where $\lambda$ is a wavelength \cite{alu}. 
Below we employ this fact to elucidate
the effect produced by a nanoparticle composed of a dielectric spherical core and a homogenous 
metallic shell on filtering phenomena at nanoscale. 

In the dipole approximation the extinction cross section 
$\sigma_{\it ext}=\sigma_{\it sc}+\sigma_{\it ab}$  is
defined by means of the scattering $\sigma_{\it sc}$ 
and absorption $\sigma_{\it ab}$ ones:   
\bea
\label{sc}
\sigma_{\it sc}=\frac{8\pi}{3}k^4|\alpha|^2\;,\\
\label{ab}
\sigma_{\it ab}=4\pi k Im(\alpha)\;.
\eea
Here, $k=2\pi\sqrt{\varepsilon_m}f/c$ is a wave number and $f$ is 
a frequency of incident photon; $\varepsilon_m$ is 
a permittivity of  a surrounding medium, 
and a dipole polarizability reads as (cf \cite{boh}) 
\be
\label{pol}
\alpha=a_s^3\frac{
(\varepsilon_s-\varepsilon_m)(\varepsilon_c+2\varepsilon_s)+
\beta (\varepsilon_m+2\varepsilon_s)(\varepsilon_c-\varepsilon_s)}
{(\varepsilon_s+2\varepsilon_m)(\varepsilon_c+2\varepsilon_s)+
2\beta(\varepsilon_s-\varepsilon_m)(\varepsilon_c-\varepsilon_s)}\;,
\ee
where $\varepsilon_c$, $\varepsilon_s$ is  permittivities 
of the core and the shell, respectively;  $\beta=(a_c/a_s)^3$ and
$a_c$, $a_s$ are the core and the shell radiuses, respectively.
Hereafter, for the sake of discussion we present results for the cross sections 
in conventional units (c.u.).

Note that the permittivities are complex, in general.
\begin{figure}
\includegraphics[width=9cm]{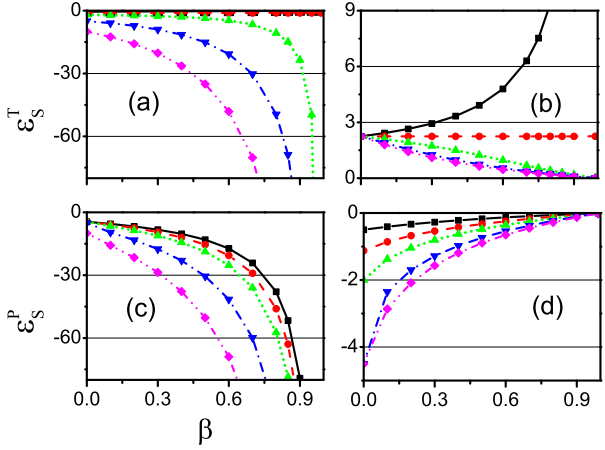}
\caption{(Color online)
Shell permittivity as a function of the ratio 
$\beta=(a_c/a_s)^3$ for various values of 
the core permittivity. Different core permittivities  are indicated by symbols:
squares connected by solid line are used for $\varepsilon_c=1$; 
circles connected by dashed line are used for $\varepsilon_c=2.25$;
up triangles connected by dotted line are used for $\varepsilon_c=4$;
down triangles connected by dash-dotted line are used for $\varepsilon_c=10$;
diamonds connected by dash-two dotted line are used for $\varepsilon_c=20$. 
The permittivity of the surrounding was set to be 2.25. 
The top panels display shell permittivity for negative (a) and positive (b) 
solutions of Eq.(\ref{st}), which provide {\it the transparency conditions}. 
The bottom panels (c), (d), display  shell permittivity for the 
solutions of Eq.(\ref{pl}), which lead to {\it the plasmon resonance}.
}
\label{fig1}
\end{figure}
It is instructive, however,  to consider first only real permittivities (no absorption). 
We assume that nanoparticles are embedded randomly in typical optical plastics 
(with the permittivity $\varepsilon_m\approx2.25$). We do not discuss a particular
device which is beyond the scope of the present paper but rather focus on
the filtering effect produced by a small nanoparticle with a metallic nanoshell.

Evidently, if the polarization is zero $\alpha=0$,
the scattering cross section of nanoparticle becomes 
zero as well. It results in the drastic increase of the transparency of 
a medium composed of such nanoparticles. 

The requirement $\alpha=0$ defines 
an analytical dependence of the shell properties on the size and dielectric 
characteristics of the nanoparticle core as
\be
\label{st}
\varepsilon_s^{T(1,2)}=-\frac{\varepsilon_m}{y}\bigg\{
\left(
t-\frac{x}{2}\right)
\pm\bigg[\left(
t-\frac{x}{2}\right)^2+
\frac{y^2}{2}
t\bigg{]}^{1/2}
\bigg\}\;,
\ee
where we have introduced the notations: $t=\varepsilon_c/\varepsilon_m$,
$x=2(\beta+2)/(2\beta+1)$, $y=4(1-\beta)/(2\beta+1)$. Thus, the orientation of
a polarization vector in the dielectric core ($\varepsilon_c>0$) is opposite to 
a local polarization vector in the nanoshell cover. As a result, a cancellation of 
the scattering caused by the nanoparticle may occur. The first solution of Eq.(\ref{st}) 
(being always negative) displays a strong dependence on 
the ratio $\beta$ (see Fig.\ref{fig1}(a)). The second solution of Eq.(\ref{st}) is 
always positive and, thus, another possibility to have a zero polarizability 
occurs. As follows from Fig.\ref{fig1}(b), the nanoparticle should be invisible at: 
i)$\varepsilon_m=\varepsilon_c=\varepsilon_s$; ii)
$\varepsilon_s>\varepsilon_m>\varepsilon_c$; 
iii)$\varepsilon_c>\varepsilon_m>\varepsilon_s$.
In other words, either the core or the shell acts as a void with the effective 
"negative" dielectric permittivity. This solution displays a weak dependence on 
the ratio $\beta$ (see Fig.\ref{fig1}(b)).

Once the denominator becomes equal to zero, the conditions are 
realized for the surface plasmons to be excited by incident light 
(an electromagnetic wave with a wave frequency $\omega=2\pi f$).
This process depends on the dielectric constant of the metal's surface, 
and this effect is exploited in surface plasmon 
resonance spectroscopy. For nanoparticles  with the radius 
on the order of the plasmon resonance wavelength, the surface plasmon dominates 
the electromagnetic response of the structure. 

In fact, two plasmon  frequencies are produced by  the inner and 
outer surfaces of the shell (cf \cite{pr1}). Indeed, these resonances 
are excited, if the shell permittivity is subject to the condition
\be
\label{pl}
\varepsilon_s^{P(1,2)}=-\frac{\varepsilon_m}{y}\bigg\{
\left(
t+x\right)
\pm\bigg[\left(
t+x\right)^2-
y^2
t\bigg{]}^{1/2}
\bigg\}\;.
\ee
In the Drude approximation (without absorption) 
$(\omega_{P}/\omega)^2=1-\varepsilon_s^{P}(\omega)$, where
$\omega_P$ is a plasmon frequency of the bulk metal;
and two solutions for the shell permittivity provide two plasmon frequencies.
The plasmon frequencies can be varied widely with 
the variation of the inner to outer shell radius ratio
(or the ratio $\beta$) (see Fig.\ref{fig1}(c)).
Evidently, the scattering cross section (\ref{sc}) is greatly enhanced.

The above analysis implies that it is possible to select such a set
of parameters which may bring together two essential ingredients,
 {\it transparency and absorption} of the complex nanoparticle,
to produce a desired filtering effect. It seems that 
for a visible and infra-red optical spectra
metals with a large negative permittivity ($Re(\varepsilon_s)\leq -10)$ and
a small absorption $Im(\varepsilon_s)$ (close to zero) would provide the best
fit to the above requirement. 
According to available sources (cf \cite{web}),
silver, gold, copper and, probably, aluminum are possible candidates for 
metallic nanoshells. Figure \ref{fig2} demonstrates that simple formulas 
Eqs.(\ref{st}),(\ref{pl}) provide a reliable evaluation of the maximal and 
minimal absorption, calculated rigorously with the aid of Eqs.(\ref{sc}),(\ref{ab}),
and (\ref{pol}); when the measured values $Re(\varepsilon_s)$, $Im(\varepsilon_s)$
are considered \cite{web}.
For example, for a spherical nanoparticle with the gold nanoshell 
(see Fig.\ref{fig2}(a) and the figure caption for the parameters) the plasmon 
oscillations, as follows from Eq.(\ref{pl}), should occur at 
$\varepsilon_s^{(1,2)}=-35,-0.5$. Taking into account the measured values 
$Im(\varepsilon_s)$ \cite{web} for this calculated $Re(\varepsilon_s)$ values, we
obtain that the first solution predicts the onset of the plasmon resonance at 
the incident light wavelength $\lambda\approx 890$ nm, 	while for 
the second solution it should occur at $\lambda\approx 355$ nm.
From rigorous results, based on the tabulated values for 
$Re(\varepsilon_c)$, $Im(\varepsilon_c)$ for different wavelengths,
one observes a strong extinction maximum at the first predicted solution, 
although  there is only the extinction growth for the second solution. 
The disagreement for the second solution may be explained  due to a relatively 
large absorption ($Im(\varepsilon_s)=5.5$ at $\lambda\approx 355$ nm) in the bare gold 
in contrast to the solution at $\lambda\approx 890$ nm, where 
$Im(\varepsilon_s)\approx 0.9$. As follows from Eq.(\ref{st}), the transparency 
for this nanoparticle should occur at $\varepsilon _{s}^{(1,2)}=-14,0.5$. 
The first solution corresponds to the incident light wavelength $\lambda\approx 680$ nm, 
where there is the extinction minimum (see Fig.{\ref{fig2}(a)), 
defined by a small absorption, indeed. 
The permittivity value corresponding to the second solution for the transparency 
is not observed for the bare gold in all examined frequency range.
According to our calculations, such nanoparticles reduce the scattering
essentially in the optical window $500< \lambda <800$ (nm), where 
$\sigma_{\it ext}\approx 340$. The ultra narrow filtering can 
occur at $\lambda \simeq 680$ nm, where the extinction cross section goes down to
$\sigma_{\it ext}\approx 50$.

\begin{figure}
\includegraphics[width=9cm]{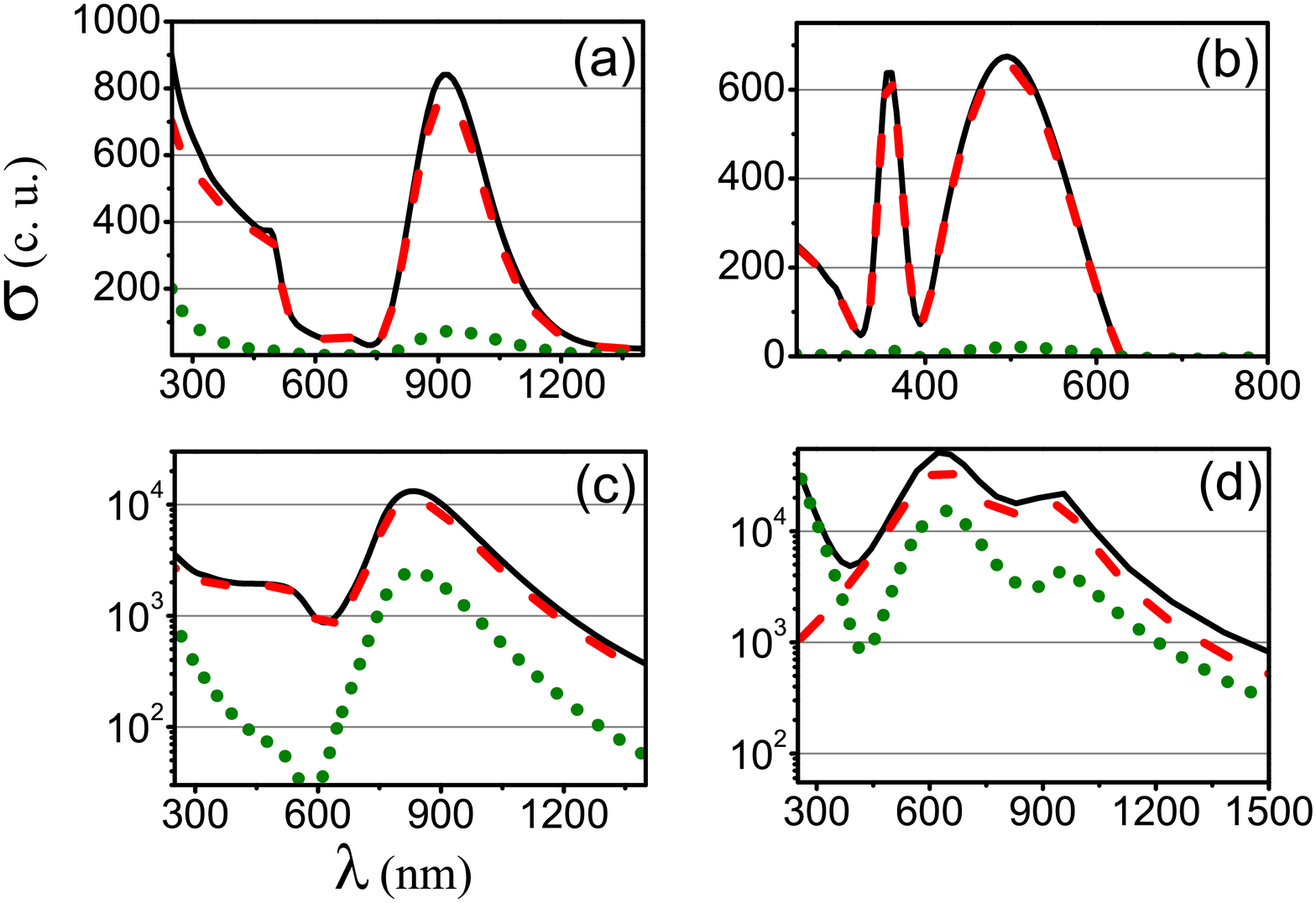}
\caption{(Color online)
Extinction (solid line), scattering (dotted line), and absorption (dashed line)
spectra as a function of the incident light wavelength for various
metallic nanoshells.
The surrounding permittivity is $\varepsilon_m=2.25$.
The nanoparticle composition : (a) gold, $\varepsilon_c=7.5$, $\beta=0.6$, 
$a_s=20$ nm; (b) silver, $\varepsilon_c=6.5$, $\beta=0.12$, 
$a_s=10$ nm; (c) cooper, $\varepsilon_c=4$, $\beta=0.62$, 
$a_s=30$ nm;  (d) aluminum, $\varepsilon_c=8$, $\beta=0.75$, 
$a_s=50$ nm; 
}
\label{fig2}
\end{figure}

For nanoparticles with the silver nanoshell the agreement between 
the simplified evaluation and rigorous results (see Fig.\ref{fig2}(b); 
the parameters are given in the figure caption) is even better. The simplified 
estimations predict the onset of the plasmon resonances at 
$\varepsilon_s^{(1,2)}=-8,-1.8$, which correspond to the wavelength 
$\lambda=490,355$ (nm), respectively \cite{web}. 
Indeed, the maxima of the extinction occur at these values 
(see Fig.\ref{fig2}(b)). As follows from Eq.(\ref{pl}), the maxima of the 
transparency are  expected at $\varepsilon_s^{(1,2)}=-3.8,1.5$. 
The first solution ($Re(\varepsilon_s^{(1)})=-3.8$ with  a small absorption 
$Im(\varepsilon_s^{(1)})\approx 0.9$ for a bare silver \cite{web})
corresponds to $\lambda=380$ nm, where the extinction has a minimum, indeed.  
From Ref.\onlinecite{web} it follows that for 
the second solution $Re(\varepsilon_s^{(2)})=1.5$ there are two 
wavelengths $\lambda=318,280$ (nm). Therefore, the second solution can be 
associated with with two values $Im(\varepsilon_s^{(2)})\approx 0.9,4$, 
respectively \cite{web}. 
For the second solution $Re(\varepsilon_s^{(2)})=1.5$ with a smaller absorption 
there is a minimum of the extinction, while for a larger value 
$Im(\varepsilon_s^{(2)})\approx4$ the rigorous results do not display 
the minimum (see Fig.\ref{fig2}(b)).
The simplified evaluation fails for the later case due to a large absorption
value $Im(\varepsilon_s^{(2)})$. 
Nanoparticles with the silver nanoshell (for a considered set of parameters) 
produce a narrow filtering effect at $\lambda=327$ nm and $\lambda=382$ nm.

According to Eq.(\ref{pl}), for nanoparticles with the
copper nanoshell 
the plasmon resonance should be expected at  $\varepsilon_s^{(1,2)}=-27,-0.3$. 
Indeed, there is a maximum of the extinction at the wavelength $\lambda=820$ nm 
($Re(\varepsilon_s^{(1)})=-27$, $Im(\varepsilon_s^{(1)})=2.7$ 
for the bare copper \cite{web})
(see Fig.\ref{fig2}(c) and the corresponding set of the parameters). There are no 
available data ($Re(\varepsilon_s^{(2)})$, $Im(\varepsilon_s^{(2)})$) for the 
second simplified solution. The transparency should be expected at 
$\varepsilon_s^{(1,2)}=-5,0.5$. The first solution corresponds to 
$\lambda=480$ nm ($Re(\varepsilon_s^{(1)})=-5$, $Im(\varepsilon_s^{(1)})=5.8$). 
The numerical results display the minimum for the extinction 
at $\lambda=620$ nm. The violation of the condition 
$|Re(\varepsilon_s^{(1)}|\gg Im(\varepsilon_s^{(1)})$ in the simplified evaluation
explains the disagreement with the rigorous results. The second solution for the permittivity for 
the transparency is not observed for the bare copper in all examined frequency range. 
Thus, nanoparticles with the copper nanoshell 
(for a considered set of parameters) produce
a narrow filtering effect at $\lambda=620$ nm.

\begin{figure}
\includegraphics[width=5cm]{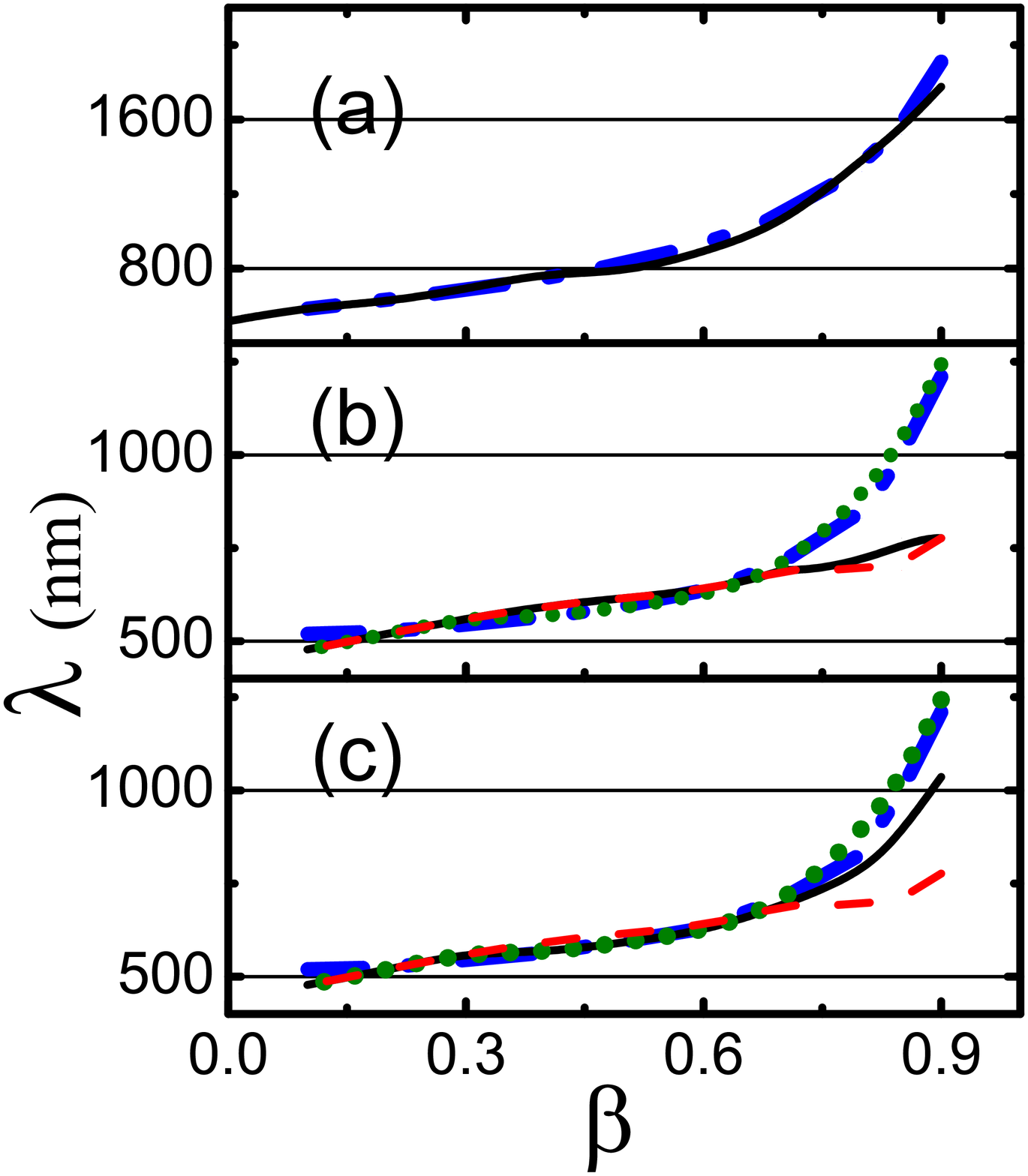}
\caption{(Color online) 
The wavelength for  the maximum and the minimum of 
the extinction cross section for nanoparticle with
the gold nanoshell $(\varepsilon_c=6.5$) as a function of the
ratio $\beta$. 
(a) At the maximum of the extinction cross section for $a_c=20$ nm:
the results based on Eq.(\ref{st}) (dash-dotted line);
the rigorous results (solid line). 
(b) At the minimum of the extinction cross section for $a_c=20$ nm:
the results based on Eq.(\ref{pl}) (dash-dotted line); 
the rigorous results for the extinction (solid line), for the scattering (dotted line),
and for the absorption (dashed line) cross sections.
(c) Similar to (b) for  $a_c=60$ nm.
}
\label{fig3}
\end{figure}

For nanoparticles with the aluminum nanoshell the large absorption is expected
from Eq.(\ref{pl}) at  $\varepsilon_s^{(1,2)}=-64.5,-0.3$ 
(see Fig.\ref{fig2}(d) and the corresponding
set of parameters). There are two wavelengths $\lambda=690,890$ (nm) 
for the absorption permittivity $Re(\varepsilon_s^{(1)})=-65.5$ which
yield two values  $Im(\varepsilon_s^{(1)})=28.6,36.7$ 
for the bare aluminum \cite{web}, respectively. There is a reasonable agreement with 
the rigorous results which display 
two maxima for the extinction  at $\lambda=620,950$ (nm) (see Fig.\ref{fig2}(d)). 
The value $\varepsilon_s^{(2)}=-0.3$ is outside of the considered optical spectra.
Eq.(\ref{st}) predicts that 
the reduced scattering is expected at $\varepsilon_s^{(1,2)}=-28,0.3$. 
According to Ref.\onlinecite{web}, the permittivituy $Re(\varepsilon_s)=-28$ 
at $\lambda= 440$ nm is accompanied by $Im(\varepsilon_s)=4.3$. 
The rigorous results display  the minimum of the
extinction at $\lambda=390$ nm which is in a reasonable agreement with the simplified
estimation. The value $\varepsilon_s^{(2)}=0.3$ is outside of the considered 
optical spectra. Thus, for nanoparticles with the  aluminum nanoshell 
(for a considered set of parameters) 
a narrow filtering occurs at $\lambda=390$ nm.

\begin{figure}
\includegraphics[width=7cm]{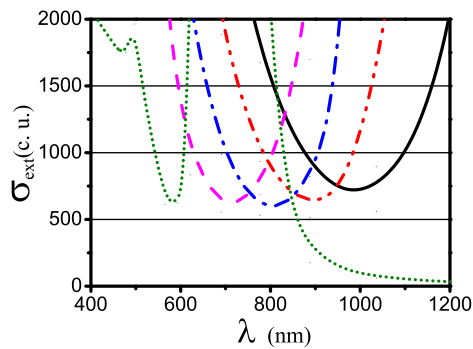}
\caption{(Color online) 
The extinction cross section for nanoparticle  with the gold nanoshell
and various cores. The following parameters are used:
$\varepsilon_c=10.6$, $\beta=0.8$, $a_s=60$ nm for ZnGeP$_2$ (solid line); 
$\varepsilon_c=8.25$, $\beta=0.8$, $a_s=60$ nm for TiO$_2$ 
(dash-double dotted line);
$\varepsilon_c=6.5$, $\beta=0.8$, $a_s=60$ nm for CdS (dash-dotted line);
$\varepsilon_c=5$, $\beta=0.8$, $a_s=60$ nm for $\alpha$GaN (dashed line);  
$\varepsilon_c=8.25$, $\beta=0.3$, $a_s=25$ nm for TiO$_2$ (dotted line). 
}
\label{fig4}
\end{figure}
The considered cases demonstrate the importance of the absorption for the 
extinction cross section. The extinction maxima as a function of the wavelength 
are well described by a simple expression (\ref{pl}) (see Fig.\ref{fig3}). 
The analytical estimation (\ref{st}) is most efficient for small nanoparticles 
$(a_s\sim 20$ nm) with relatively small ratio core to shell radius, albeit it 
provides a reasonable agreement for larger nanoparticles $(a_s\sim 60$ nm) at 
large ratios $\beta$.
Evidently, the location of the extinction maxima and minima can be varied 
by a proper choice of the core material as well. Figure 4 displays the minima 
of the extinction for various semiconductors. 
For example, in the optical window $450<\lambda<630$ (nm) 
a narrow filtering takes place at  $\lambda\approx 565$ nm 
for small nanoparticles ($a_s=20$ nm) with the gold 
nanoshell and the core TiO$_2$ (see Fig.\ref{fig4}). In this case the 
extinction cross section decreasing by three times from the border sets 
to the center of the window.

In conclusion, we propose and analyze in depth filtering with the aid of 
the plasmonic shell surrounding the dielectric core.
To this end we focussed on a plasmonic nanoshell made from
different metals. It was demonstrated that such a nanoshell allows:
i)to reduce the scattering and the extinction cross sections of the complex nanoparticle; 
and ii)to increase the absorption, -- by several times beyond the desired optical window.
Note that the fabrication of such a metallic nanoshell is amenable by self-organization 
methods in colloidal nanochemistry.

\section*{Acknowledgement}
This work is partly supported by Russian Federal Program 
Grant No 2011-1.3-513-054-006, RFBR Grant No.11-02-00086
(Russia), and Spanish MICINN Grant No. FIS2008-00781.

\end{document}